\documentclass{epl}

\title{One-Bead Microrheology with Rotating Particles}
\shorttitle{One-Bead Microrheology with Rotating Particles}
\author{M. Schmiedeberg \and H. Stark \thanks{E-mail: 
\email{Holger.Stark@uni-konstanz.de}}}
\shortauthor{M. Schmiedeberg \and H. Stark}
\institute{Universit\"at Konstanz, Fachbereich Physik, D-78457
Konstanz, Germany}

\begin{document}

\maketitle

\begin{abstract}
We lay the theoretical basis for one-bead microrheology with rotating
particles, i.e, a method where colloids are used to probe the mechanical
properties of viscoelastic media. Based on a two-fluid model, we calculate
the compliance and discuss it for two cases. We first assume that the
elastic and fluid component exhibit both stick boundary conditions at the
particle surface. Then, the compliance fulfills a generalized Stokes law 
with a complex shear modulus whose validity is only limited by inertial
effects, in contrast to translational motion. Secondly, we find that
the validity of the Stokes regime is reduced when the elastic network
is not coupled to the particle.
\end{abstract}

\section{Introduction}
In recent years the experimental method of microrheology\ \cite{microrheo}
emerged as a powerful tool to monitor the mechanical properties of 
viscoelastic soft materials\ \cite{viscoelastic1,Ferry1980} especially in 
biological systems\ \cite{viscoelastic2,Schnurr1997} including cells\ 
\cite{Bausch1999}. 
The main idea is to disperse micron-sized beads into the material
and monitor their motion either as reponse to external forces
(active method)\ \cite{Bausch1999,active} or due to Brownian fluctuations 
(passive method)\ \cite{Schnurr1997,passive}. Whereas in the first method, the 
frequency-dependent response function or compliance is measured directly, 
it is inferred in the second method from the particle's positional 
fluctuations using the fluctuation-dissipation theorem and the
Kramers-Kronig relations\ \cite{Schnurr1997}.

In this article, we address the particles' rotational degree of freedom
and lay the theoretical basis to use it in microrheology. To monitor
the particles' orientations, they have to be anisotropic. And indeed,
first experimental implementations of one-bead microrheology with 
either birefringent spherial particles\ \cite{rotating} or microdisks\ 
\cite{Cheng2003} do exist.

The main theoretical problem of microrheology is how the particle 
compliance measured in experiments relates to the viscoelastic properties 
of the material quantified in its complex shear modulus. We will show here,
based on a model-viscoelastic medium, the so-called 
two-fluid system\ \cite{twofluid},
that the compliance in the rotational case obeys a generalized Stokes 
law which is only limited by an upper crossover frequency in contrast
to the translational motion where the validity is restricted to a frequency 
window\ \cite{Schnurr1997,Levine}. 
So one advantage of rotating particles is that they extend the 
frequency range that allows a straightforward interpretation 
of the experimental results by the Stokes law.

Let us first review the reason for the frequency window in the 
translational case\ \cite{Schnurr1997}. As a model-viscoelastic medium, 
we consider an elastic component, e.g., a polymer network, which can move 
relative to a Newtonian fluid. A particle embedded in this medium experiences
an external oscillating force with amplitude $F(\omega)$ 
and reacts with an oscillating displacement described by the amplitude 
$x(\omega) = \alpha(\omega) F(\omega)$, where $\alpha(\omega)$ is the
compliance as a function of frequency $\omega$. For $\omega=0$, the 
displaced particle creates a deformation field in the elastic network 
which includes both shear and compressional contributions. For small 
$\omega$, the elastic network thus has to move relative to the 
incompressible fluid since the latter only allows shear motions. However, the 
frictional force between both components increases with their relative 
speed. Hence, beyond a crossover frequency $\omega_{c1}$, elastic
network and viscous fluid are strongly coupled and can be considered as a
single incompressible viscoelastic medium. The compliance for a spherical
particle then assumes the simple generalized Stokes relation 
$\alpha(\omega) = [6\pi G(\omega)a ]^{-1}$ where 
$G(\omega)=\mu-i\omega\eta$ is the complex shear modulus, $\eta$ is 
viscosity and $\mu$ denotes Lam\'e's elastic constant associated with 
shear. Beyond a second crossover frequency $\omega_{c2}$, inertial effects 
of the fluid become important and the generalized Stokes relation is no 
longer valid. This scenario is confirmed by detailed calculations using
a ``volume localization'' approximation\ \cite{Levine}.

\section{Theory}
In the following, we consider a rotating particle with an oscillating 
angular displacement $\bm{\Phi}(t) = \bm{\Phi}(\omega) e^{-i\omega t}$ 
where the direction of the vector $\bm{\Phi}(\omega)$ caracterizes the 
axis of rotation. Since the pressure around the particle stays constant,
the oscillating velocity field in a pure incompressible Newtonian fluid is 
described
by the Helmholtz equation $(\nabla^{2}+k^{2})\bm{v}(\bm{r},\omega)=0$ with 
wave number $k=\sqrt{i\omega\rho/\eta}$\ \cite{Landau1991}. 
Under stick boundary conditions it assumes the form
\begin{equation}
\bm{v}(\bm{r},\omega) = -i\omega \left(\frac{a}{r}\right)^{3} 
[\bm{\Phi}(\omega) \times \bm{r}] \frac{1-ikr}{1-ika}e^{ik(r-a)} \enspace
\end{equation}
which determines the external torque 
$T(\omega) = \alpha^{-1}(\omega)\Phi(\omega)$ to drive the oscillating
particle. For small frequencies, the compliance obeys the familiar 
Stokes result, $\alpha^{-1}/-i\omega = 8\pi a^{3}\eta$. Deviations from
this law, occur around $\omega_{0}=2\eta/(\rho a^{2})$, i.e., when the 
penetration depth $\delta = \mathrm{Im}(k)$ equals the particle radius $a$.

We now study the equivalent problem for a model viscoelastic medium,
represented by a two-fluid model\ \cite{twofluid}:
\begin{eqnarray}
\bm{0} & = & \mu \nabla^{2} \bm{u} + (\lambda+\mu) \nabla(\nabla \cdot \bm{u})
+ \Gamma (\bm{v}-\frac{\partial \bm{u}}{\partial t}) \label{2.1}\\
\rho\frac{\partial}{\partial t} \bm{v} & = & -\nabla p + \eta \nabla^{2} \bm{v}
- \Gamma (\bm{v}-\frac{\partial \bm{u}}{\partial t}) \enspace , \enspace
\mathrm{div} \bm{v}=0 \enspace. \label{2.2}
\end{eqnarray}
Here
an incompressible Newtonian fluid with shear viscosity $\eta$ 
is coupled to an elastic medium with Lam\'e constants $\lambda$, $\mu$ 
via a conventional friction term.
By dimensional analysis, the friction coefficient 
$\Gamma = \eta / \xi^{2}$ contains a characteristic length $\xi$ which is
on the order of the mesh size in, e.g., an actin network\ \cite{Levine}. 
We neglect the mass density of the elastic medium to the one of the 
Newtonian fluid right from the beginning. The characteristic parameters
of the theory are the reduced mesh size $\xi/a$, 
$\omega_{e}=\mu/\eta$ and $\omega_{0} = 2 \eta/(\rho a^{2})$.
The two frequencies quantify the fluid inertia, shear elasticity, and
shear viscosity. Typical numbers for an actin solution\ \cite{Schnurr1997} 
are $\omega_{e} = 10^{3}\,\mathrm{Hz}$ and
$\omega_{0}=10^{5}\,\mathrm{Hz}$ based on $a=3\,\mathrm{\mu m}$, 
$\eta = 0.01 P$ and $\mu=1\,\mathrm{N/m^{2}}$.

An oscillating rotating particle creates pure shear fields for both 
displacement $\bm{u}$ and velocity $\bm{v}$ 
($\mathrm{div} \bm{u} = \mathrm{div} \bm{v}=  0$) and keeps the pressure
constant. One then derives from Eqs.\ (\ref{2.1}) and (\ref{2.2}) that
the amplitudes $\bm{u}(\bm{r,}\omega)$ and $\bm{v}(\bm{r},\omega)$
of the oscillating fields obey a vector Helmholtz equation in analogy
to the pure Newtonian fluid, however with $k^{2}$ replaced by a 
matrix $\bm{K}^{2}$. To solve this equation, we introduce
\begin{equation}
\bm{u}(\bm{r},\omega) = - a^{2} \bm{\Phi}(\omega) \times \nabla g(r)
\enspace \mathrm{and} \enspace \bm{v}(\bm{r},\omega) = - \omega_{e} a^{2} 
\bm{\Phi}(\omega) \times \nabla h(r)
\end{equation}
and finally arrive at the equivalent Helmholtz equation for
$g(r)$ and $h(r)$:
\begin{equation}
(\overline{\nabla}^{2} \bm{1} + \bm{K}^{2})
   \left(\begin{array}{c} g(\bar{r}) \\ h(\bar{r}) \end{array} \right) 
= \bm{0} \enspace \mathrm{with} \enspace
\bm{K}^{2} = \frac{a^{2}}{\xi^{2}} 
   \left(\begin{array}{cc} 
           i \omega/\omega_{e} & 1 \\        
          -i \omega/\omega_{e} & 2i \omega \xi^{2} / (\omega_{0}a^{2}) - 1
   \end{array} \right) \enspace.
\label{5}
\end{equation}
The reduced radial coordinate is denoted by $\bar{r}=r/a $ and the radial
part of the Laplace operator reads
$\overline{\nabla}^{2} = \frac{1}{\bar{r}^{2}} 
\frac{\partial}{\partial \bar{r}} (\bar{r}^{2} 
\frac{\partial}{\partial \bar{r}})$. 
The general solution of Eq.\ (\ref{5}) is given by 
$\exp(i\bm{K}\bar{r})/\bar{r}$ or 
\begin{equation}
\left(\begin{array}{c} g(\bar{r}) \\ h(\bar{r}) \end{array} \right)
 = \frac{1}{\bar{r}} \, \bm{S}
   \left(\begin{array}{cc} 
           \exp(i\sqrt{\lambda_{1}}\bar{r} ) & 0\\        
           0 & \exp(i\sqrt{\lambda_{2}}\bar{r})
   \end{array} \right) \,
\left(\begin{array}{c} b_{1} \\ b_{2} \end{array} \right) \enspace,
\label{6}
\end{equation}
where $\lambda_{i}$ are the eigenvalues of $\bm{K}^{2}$. The matrix
$\bm{S} = (\bm{e}_{1} , \bm{e}_{2})$ is composed of the eigenvectors 
$\bm{e}_{1}$ and $\bm{e}_{2}$ and therefore diagonalizes $\bm{K}^{2}$.
Finally, the constants $b_{i}$ are determined by the boundary conditions
on the surface of the particle. Note that the roots $\sqrt{\lambda_{i}}$ 
have to be chosen such that the exponentials in Eq.\ (\ref{6}) decay to 
zero for large $\bar{r}$.

In general, the external torque on a particle is calculated by 
$\bm{T} = -\int \bm{r} \times \bm{\sigma} d\bm{f}$ where  $\bm{\sigma}$
is the stress tensor and $d\bm{f}$ the directed surface element. 
In our case, $d\bm{f} \propto -\bm{r}$ and the velocity $\bm{v}$ and 
displacement vector $\bm{u}$ point along the azimuthal direction 
relative to the axis $\bm{\Phi}(\omega)$. Therefore, only the 
respective components of the elastic and viscous stress tensor,
\begin{equation}
\sigma_{\varphi r}^{u} = \mu 
\Big(\frac{\partial u_{\varphi}}{\partial r} - \frac{u_{\varphi}}{r}\Big)
\quad \mathrm{and} \quad \sigma_{\varphi r}^{v} = \eta
\Big(\frac{\partial v_{\varphi}}{\partial r} - \frac{v_{\varphi}}{r}\Big)
\enspace,
\label{7}
\end{equation}
contribute and give a torque parallel to $\bm{\Phi}(\omega)$ with 
magnitude $T=T^{u} + T^{v}$, where
\begin{eqnarray}
T^{u} & = & 8\pi\mu a^{3} \Phi(\omega) 
\Big(1+\frac{i\omega}{3\omega_{e}} 
[S_{11}b_{1}\lambda_{1}e^{i\sqrt{\lambda_{1}}} +
 S_{12}b_{2}\lambda_{2}e^{i\sqrt{\lambda_{2}}}]\Big) 
\label{8} \\
T^{v} & = & -i\omega 8\pi\eta a^{3} \Phi(\omega) 
\Big(1-\frac{1}{3} 
[S_{21}b_{1}\lambda_{1}e^{i\sqrt{\lambda_{1}}} +
 S_{22}b_{2}\lambda_{2}e^{i\sqrt{\lambda_{2}}}]\Big) \enspace .
\label{9}
\end{eqnarray}
The prefactors on the right-hand side of Eqs.\ (\ref{8}) and (\ref{9})
are the Stokes results for a pure elastic or viscous medium.

In the following we consider two cases to determine the constants 
$b_{i}$ from the boundary conditions. In the first case, we assume stick
boundary conditions for both the viscous and the elastic component at the
particle surface, i.e.,
$\bm{v}=-i\omega \bm{u} = \left. -i\omega \bm{\Phi}(\omega) \times 
\bm{r}\right|_{r=a}$. So we assume that the elastic network is attached
to the particle.
In the second case, the elastic network
is not attached and the elastic stress tensor $\sigma_{\varphi r}^{u}$
in Eqs.\ (\ref{7}) vanishes 
on the particle surface. As a result, the elastic torque $T^{u}$ is zero 
and the displacement field has to fulfill the mixed boundary condition 
$\left.(\partial u_{\varphi}/\partial r -u_{\varphi}/r) \right|_{r=a} =0$.
In both cases we can write down the solutions in
analytic form. However, the concrete formulas are too large to convey
direct information. We therefore discuss their graphic representations 
and determine certain limits.

\section{Case 1}
\begin{figure}
\twofigures[height=6.5cm]{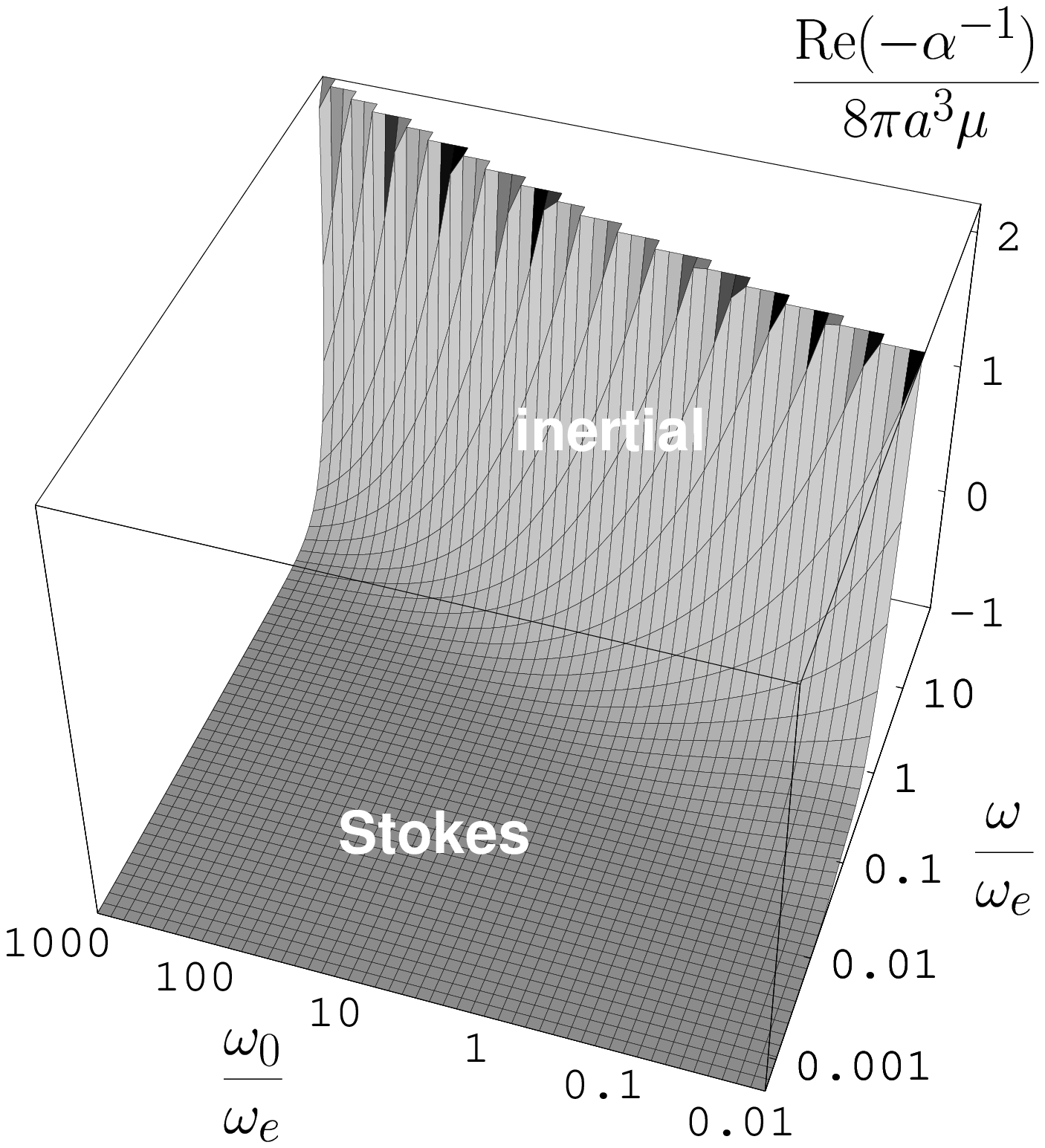}{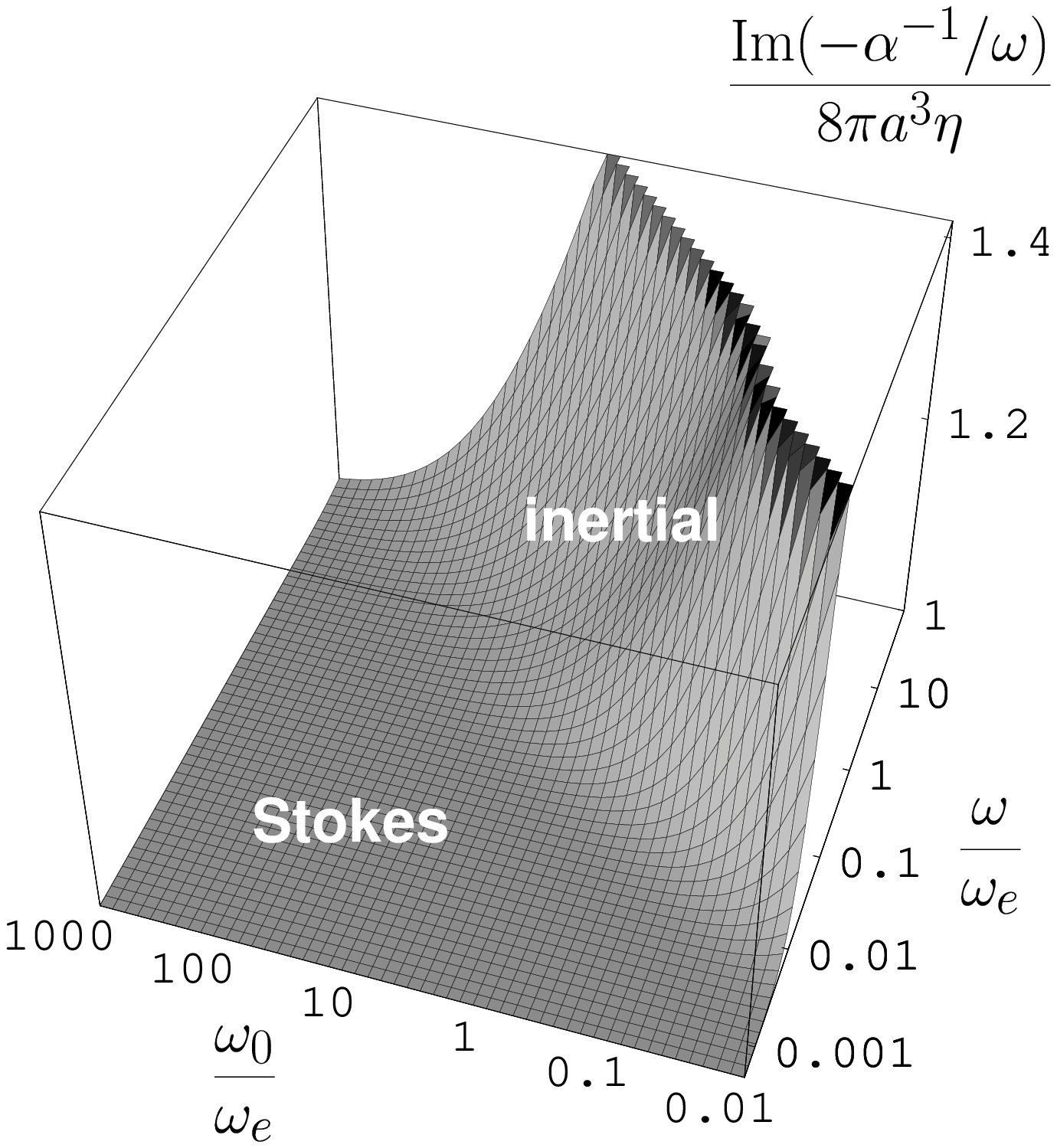}
\caption{Case 1:
Real part of the inverse compliance $\alpha^{-1}$, normalized to
$8\pi a^{3}\mu$ as a function of 
$\omega/\omega_{e}$ and $\omega_{0}/\omega_{e}$; the parameter is $\xi/a=0.1$.}
\label{f.1}
\caption{Case 1: Imaginary part of the inverse compliance $\alpha^{-1}$, 
normalized to $\omega 8\pi a^{3}\eta$ as a function of 
$\omega/\omega_{e}$ and $\omega_{0}/\omega_{e}$; the parameter is $\xi/a=0.1$.}
\label{f.2}
\end{figure}
In Figs.\ \ref{f.1} and \ref{f.2} we plot the real and imaginary part
of the inverse compliance $\alpha^{-1}(\omega)$, relative to their results 
for a pure elastic and viscous system, as a function of $\omega/\omega_{e}$
and $\omega_{0}/\omega_{e}$; the mesh size is $\xi/a =0.1$.
For low frequencies they both exhibit constant values which correspond to 
the generalized Stokes relation $\alpha^{-1} = 8\pi a^{3} G(\omega)$ 
with $ G(\omega) = \mu - i\omega \eta$. This result can be extracted
from our theory as long as the term 
$2\omega \xi^{2} / (\omega_{0}a^{2})$ in the matrix $\bm{K}^{2}$
of Eq.\ (\ref{5}) can be neglected against one, i.e., as long as
inertial effects of the fluid are negligible.
Note that unlike the translational motion, the validity of the Stokes relation
extends to $\omega \rightarrow 0$; there is no lower crossover frequency.
The reason is clearly that a rotating sphere only creates pure shear
fields for both dynamic variables $\bm{u}$ and $\bm{v}$. This indicates
that elastic network and viscous fluid are strongly coupled to each other 
and move together. Therefore, their dynamics is described by the equation 
of linear elasticity with the complex shear modulus $G(\omega)$
(reminiscent to a Voigt element\ \cite{Ferry1980})
or alternatively by the Navier-Stokes equation with $\eta$ replaced by 
$\eta - \mu /(i\omega)$. With the latter view, the compliance can be
calculated using the result from the Newtonian fluid \cite{Landau1991}
with a complex wave number given by $k^{2}=\rho \omega^{2}/(\mu-i\omega\eta)$:
\begin{equation}
\alpha^{-1}(\omega) = 8\pi a^{3} G(\omega) 
\frac{1+2 \sqrt{x}+2x+2x^{3/2}/3-i2x(1+\sqrt{x})/3}{1+2\sqrt{x}+2x}
\enspace, \enspace
x = \frac{\omega}{\omega_{0}} \frac{1}{1+i\omega_{e}/\omega} \enspace.
\label{10}
\end{equation}
The last equation fits the graphs in Figs.\ \ref{f.1} and \ref{f.2} well.
It especially accounts for the deviations from the Stokes law due to
inertial effects at higher frequencies. In a pure Newtonian fluid, 
inertia becomes noticable around the frequency $\omega_{0}$; 
just set $\omega_{e}=\mu/\eta=0$ in Eq.\ (\ref{10}). For the strongly coupled
viscoelastic system, the onset of inertial effects is not so clear
from Eq.\ (\ref{10}). We therefore determined the appropriate crossover
frequencies empirically by requiring that the compliance $\alpha(\omega)$
deviates from the  Stokes law by 10\%. As it is already obvious from the
graphs in Figs.\ \ref{f.1} and \ref{f.2}, we find that the 
crossover frequencies exhibit different behavior for the real and imaginary 
part of $\alpha^{-1}(\omega)$. For the real part, it scales as
$\sqrt{\omega_{0}\omega_{e}}\propto \mu/(\rho a^{2})$ whereas for the
imaginary part it behaves as $\omega_{0}^{0.77} \omega_{e}^{0.23}$ for
$\omega_{0}/\omega_{e}<1$ and passes over to $\omega_{0}$ for
$\omega_{0}/\omega_{e}>1$. This cross over is clearly seen in Fig.\ \ref{f.2}.

So far, we discussed the case of $\xi/a \ll 1$.
For $\xi/a \sim 1$, the regime of generalized Stokes law still exists but
Eq.\ (\ref{10}) does not apply anymore, although the deviations are not
dramatic. The regime $\xi/a \sim 1$ means that the continuum limit can
no longer be applied to elastic networks such as actin. However, in
the two-fluid model $\xi$ just quantifies the frictional coupling 
between fluid and elastic component. One could wonder if there exists
a viscoelastic system with such a weak coupling so that $\xi/a \sim 1$ 
is applicable.

\section{Case 2}
\begin{figure}
\twofigures[height=6.5cm]{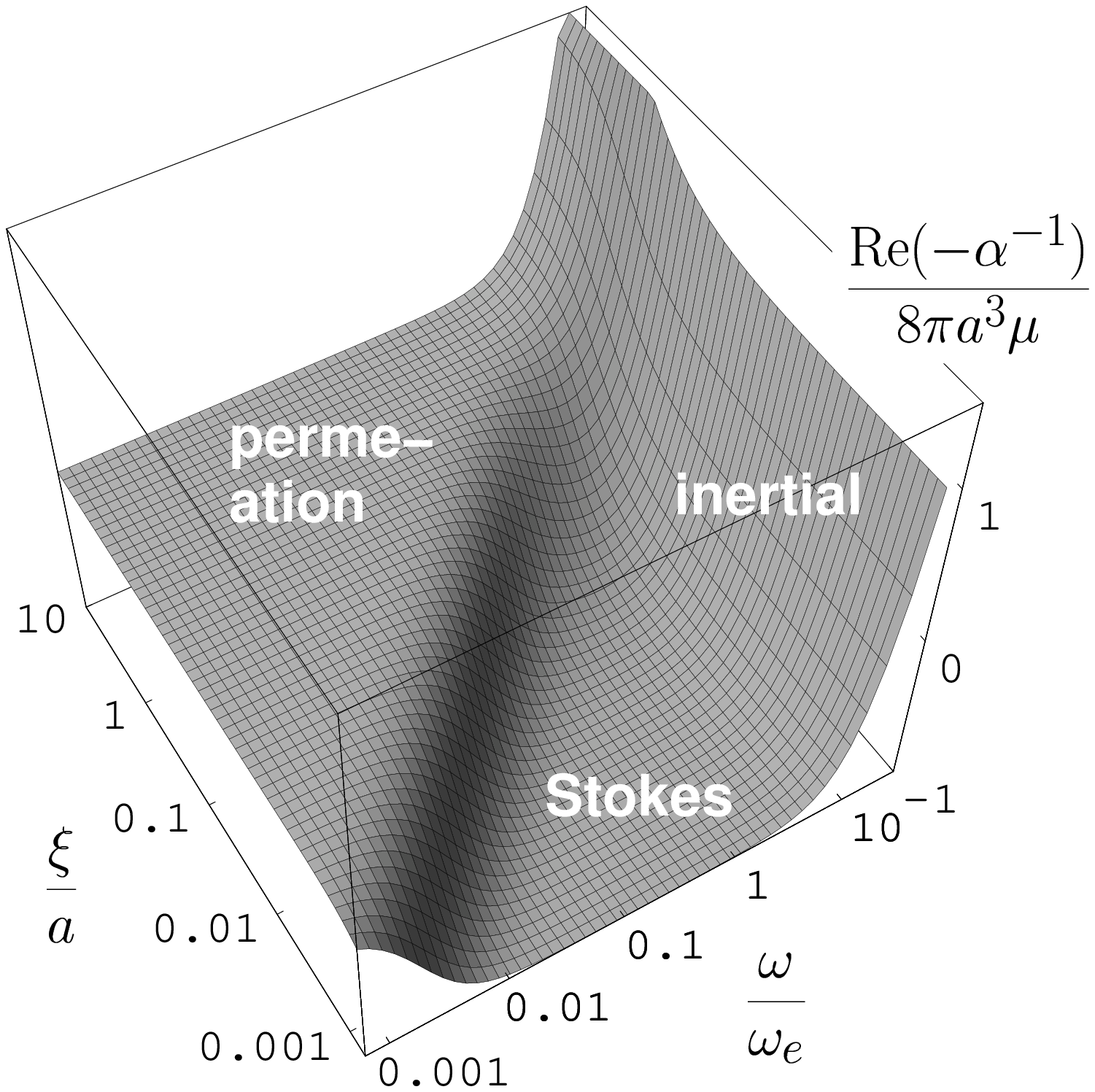}{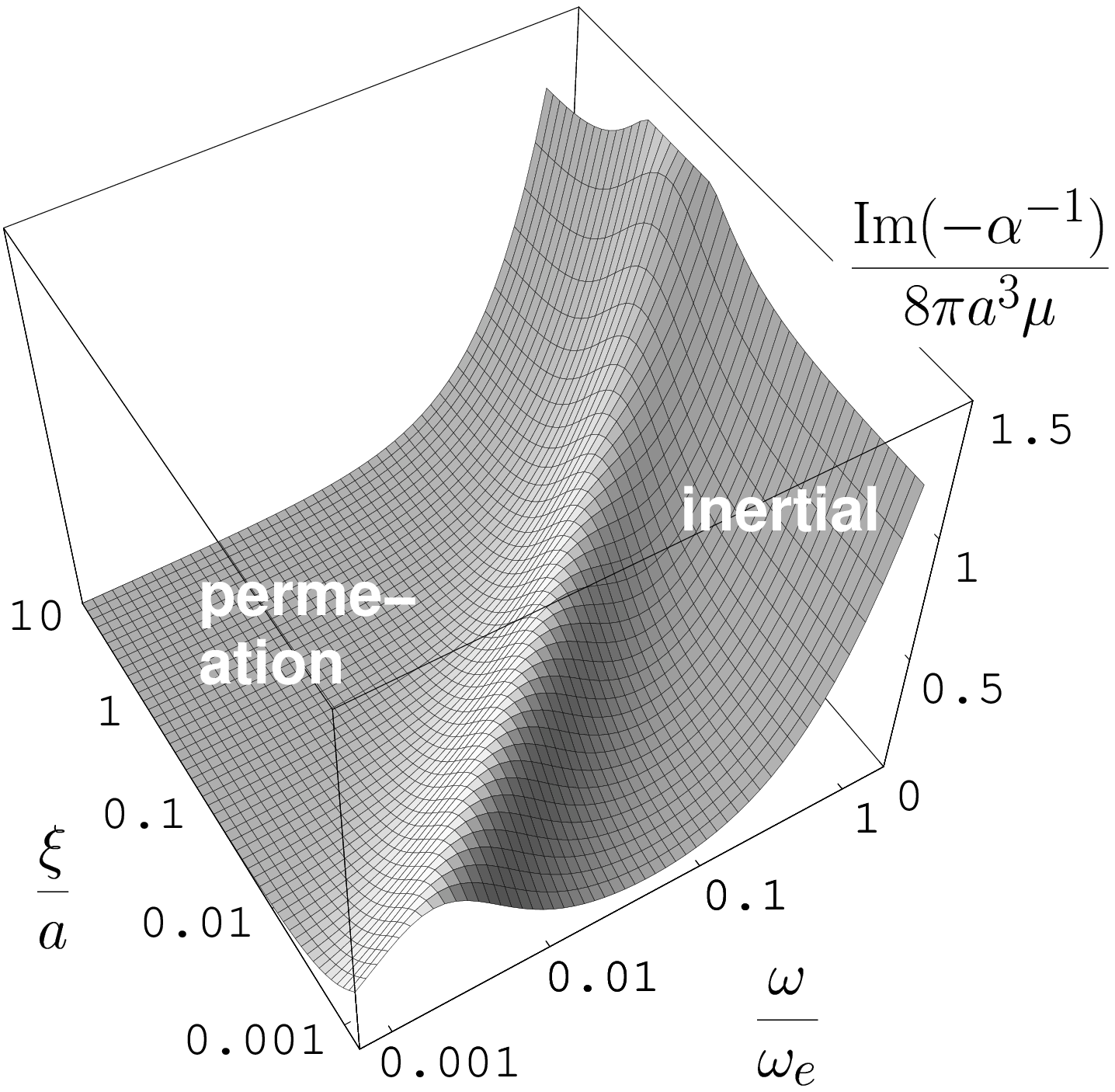}
\caption{Case 2: Real part of the inverse compliance $\alpha^{-1}$, 
normalized to $8\pi a^{3}\mu$ as a function of $\omega/\omega_{e}$ and 
$\xi/a$; the parameter is $\omega_{0}/\omega_{e} = 100$.}
\label{f.3}
\caption{Case 2: Imaginary part of the inverse compliance $\alpha^{-1}$, 
normalized to $8\pi a^{3}\mu$ as a function of $\omega/\omega_{e}$ and 
$\xi/a$; the parameter is $\omega_{0}/\omega_{e} = 100$}
\label{f.4}
\end{figure}
We now address the case where the elastic network is not coupled to the 
particle surface, it only reacts to shear flow via the friction term 
in Eq.\ (\ref{2.1}). In Fig.\ \ref{f.3}, we plot the real part of the
inverse compliance as a function of reduced frequency $\omega/\omega_{e}$ 
and mesh size $\xi/a$; the additional parameter is set to 
$\omega_{0}/\omega_{e}=100 $. For small frequencies, the real part is 
close to zero, in contrast to case 1, where it assumes the reference Stokes 
value of $8\pi a^{3}\mu$, as already discussed.
The friction between the two components is sufficiently small so the fluid 
permeates the elastic network without deforming it noticeably.
Then for increasing frequency, an edge occurs and
$\mathrm{Re}(\alpha^{-1})$ enters the region where the Stokes law is valid.
Note, however, that this region is considerably reduced compared to case 1.
Correspondingly, the imaginary part in Fig.\ \ref{f.4} (also in units of 
$8\pi a^{3}\mu$!) exhibits a ridge. This is what we roughly expect since
real and imaginary part of the compliance are connected via Kramers-Kronig 
relations\ \cite{Chaikin1995}. 
The features described so far can be explained by the formula
\begin{equation}
\alpha^{-1}(\omega) = 8\pi \mu a^{3} \left[
\frac{\omega^{2}\tau^{2}}{1-i\omega \tau} -i\omega \tau 
\Big(1- \frac{1}{\omega_{e}\tau}\Big) \right] \quad \mathrm{with}
\quad \tau = \frac{1}{\omega_{e}}\frac{a}{3\xi}\sqrt{1-i\omega/\omega_{e}}
\enspace,
\label{11}
\end{equation}
which follows from our theory when we neglect again the ``inertial term" 
$2\omega \xi^{2} / (\omega_{0}a^{2})$ in the matrix $\bm{K}^{2}$
of Eq.\ (\ref{5}) and set $\xi/a \ll 1$. To determine $\tau$, the complex root
with positive real part has to be taken. For $\omega/\omega_{e} \ll 1$, the 
existence of the edge and ridge in relation\ (\ref{11}) are obvious.
Furthermore, the relation demonstrates that edge and ridge, or the onset of 
the Stokes regime, occur at a frequency which scales as $\xi/a$. This is
in contrast to translational motion where it scales as $(\xi/a)^{2}$\ 
\cite{Schnurr1997,Levine}.
\begin{figure}
\onefigure[height=6.5cm]{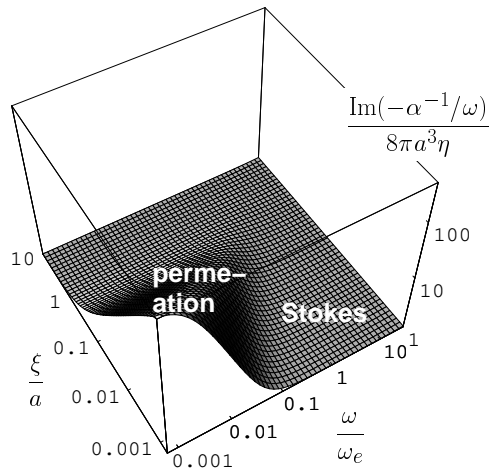}
\caption{Case 2: Imaginary part of the inverse compliance $\alpha^{-1}$, 
normalized to $\omega 8\pi a^{3}\eta$ as a function of $\omega/\omega_{e}$ 
and $\xi/a$; the parameter is $\omega_{0}/\omega_{e} = 100$}
\label{f.5}
\end{figure}
In Fig.\ \ref{f.5} the imaginary part of $\alpha^{-1}$ with proper
scale factor $\omega 8\pi a^{3}\eta$ is plotted. It reveals the strong
effect of friction between both components when the fluid permeates
the elastic network. Note the logarithmic scale of the vertical axis.
Inertial effects for $\xi/a \ll 1$ are again described by 
Eq.\ (\ref{10}). So the Stokes regime in Fig.\ \ref{f.5} extends up to 
$\omega/\omega_{e}=10$ in agreement with Fig.\ \ref{f.2}. Nevertheless,
we find that in case 2, the validity of the Stokes law is reduced.

\section{Conclusions}
Based on a two-fluid model, we determined and discussed the compliance
for a rotating spherical particle embedded in
a viscoelastic medium. In the case where both the viscous and elastic
component obey stick boundary conditions at the particle surface, we 
identify the validity of a generalized Stokes law, and therefore a 
simple relation to the complex shear modulus, starting from zero 
frequency and limited only at high frequencies by inertial effects. 
This is in contrast to translational motion. When the elastic network
is not coupled to the particle, the compliance exhibits a region
of low elasticity and high effective viscosity starting from zero frequency
indicating the strong friction which occurs when fluid permeates the elastic
network. The regime of the Stokes law is reduced.

Real viscoelastic systems such as actin solutions possess a 
frequency-depended complex shear modulus\ 
\cite{viscoelastic1,viscoelastic2,Schnurr1997}. 
So to discuss the
frequency dependence of the compliance, e.g., in case 1, one has to 
consider a non-trivial path in Figs.\ \ref{f.1} and \ref{f.2} determined
by the varying parameter 
$\omega_{0}/\omega_{e} \propto \eta^{2}(\omega)/\mu(\omega)$.
This is, however, no problem in the Stokes regime but leads to additional
effects outside the Stokes regime as, e.g., in case 2.

Our study of different boundary conditions clarifies that the 
interpretation of compliances measured in experiment needs care.
So far, in the translational case, always stick-boundary conditions
are assumed. Our results show that some slip of the elastic network
changes the measurable compliance dramatically. This could lead to
false interpretations.

Clearly, one-bead microrheology with rotating particles also suffers
the drawback that it does not probe bulk properties, as it is done
with two-bead methods\ \cite{two-bead}. On the other hand, 
since in the Stokes regime the shear fields decay as $1/r^{2}$ 
instead of $1/r$ for translational motion, 
the method could be used to monitor explicitely the effect of the 
embedded particle on the elastic network\ \cite{MacKintosh2004}.
A challenge will be to develop the theory for two-bead microrheology
including the rotational degree of freedom\ \cite{Reichert2003}.

In this article we laid the theoretical basis for microrheology with
rotating particles. Our results demonstrate that it can be a useful 
extension of existing methods based on translational motion and also 
complements them. So we hope that our work stimulates further experimental 
investigations which use rotating particles as colloidal probes to 
explore the mechanical properties of soft materials, especially in 
connection with biological systems.


\acknowledgments
The authors thank F. MacKintosh and T. Liverpool for helpful and encouraging 
discussions. They also acknowledge financial support of the 
Deutsche Forschungsgemeinschaft under Grant No. Sta 352/5-2 and within the 
International Graduate College ``Soft Matter'' and the transregio SFB 6 
``Physics of Colloidal Dispersions in External Fields".





\end{document}